\documentstyle[11pt]{article}

\def\ph{\phantom{-}}
\def\za{\phantom{0}}
\def\zb{\phantom{00}}
\def\zc{\phantom{000}}
\begin{document}

\begin{titlepage}

\vskip0.5cm

\begin{flushright}
MS-TPI-96-1 \\
KL-TH-96/2 \\
DAMTP 96-03 \\ 
January 1996
\end{flushright}
\vskip0.5cm

{\Large\bf 
\begin{center} Iterating Block Spin Transformations     \\
         of the $O(3)$ Non-Linear $\sigma$-Model \end{center}
}
\vskip1.3cm
\centerline{A.P. Gottlob$^a$, M. Hasenbusch$^b$, and K. Pinn$^{c}$}
\vskip5mm
\centerline{\sl $^a$ Fachbereich Physik, Universit\"at Kaiserslautern,}
\centerline{\sl D-67653 Kaiserslautern, Germany
\footnote{e--mail: gottlob~@physik.uni-kl.de}}
\vskip2mm 
\centerline{\sl $^b$ DAMTP, Silver Street,}
\centerline{\sl Cambridge, CB3 9EW, England
\footnote{e--mail: mh205@amtp.cam.ac.uk}}
\vskip2mm
\centerline{\sl $^c$ Institut f\"ur Theoretische Physik I, Universit\"at
                M\"unster}
\centerline{\sl Wilhelm-Klemm-Str.\ 9, D-48149 M\"unster, Germany
\footnote{e--mail: pinn@uni-muenster.de}}

\vskip2.0cm

\begin{abstract}

We study the iteration of block spin transformations in the  $O(3)$
symmetric non-linear $\sigma$-model on a two-dimensional square
lattice with help of the Monte Carlo method.
In contrast to the classical Monte Carlo Renormalization Group
approach, we {\em do} attempt to explicitly compute the  block spin
effective actions.  Using two different methods for the determination
of effective couplings, we study the
renormalization group flow for various parametrization and truncation
schemes. The largest ansatz for the effective action
contains thirteen coupling constants.

Actions on the renormalized trajectory should describe  theories
with no lattice artefacts, even at small correlation  length. However,
tests with the step scaling function  of L\"uscher et al.\ reveal that
our truncated effective  actions show sizable scaling
violations indicating that the ansaetze are still too small.  
  
\end{abstract}

\end{titlepage}
\newcommand{\nc}{\newcommand}

\nc{\be}{\begin{equation}}
\nc{\ee}{\end{equation}}
\nc{\bea}{\begin{eqnarray}}
\nc{\eea}{\end{eqnarray}}

\nc{\rbo}{\raisebox}
\nc{\cH}{{\cal H}}
\nc{\RR} {\rangle }
\nc{\LL} {\langle }
\nc{\rmi}[1]{{\mbox{\small #1}}}
\nc{\eq}[1]{eq.~(\ref{#1})}
\nc{\Eq}[1]{Eq.~(\ref{#1})}
\nc{\ul}{\underline}
\nc{\mc}{\multicolumn}
\nc{\todo}[1]{\par\noindent{\bf $\rightarrow$ #1}}
\nc{\nonu}{\nonumber}

\nc{\half}{\mbox{\small$\frac12$}}
\nc{\eights}{\mbox{\small$\frac18$}}

\nc{\va}{\varphi_x^{\alpha}}
\nc{\Lx}{{\bf L}}
\section{Introduction}

The Monte Carlo Renormalization Group (MCRG) \cite{ma,swendsen,baillie}
combines  ideas of the block spin renormalization group (RG) and Monte
Carlo (MC) simulations.

In the traditional MCRG one refrains from explicitly computing effective
actions.\footnote{What is called action in Euclidean quantum field
theory, is called Hamiltonian in classical statistical mechanics.}
Instead, one employs the RG as a tool to define blocked observables 
suitable for the efficient computation of critical properties. Examples
for these techniques are the methods for the determination  of critical
exponents from the linearized RG transformation \cite{swendsen} and the
matching method for the calculation of the $\Delta \beta$
function~\cite{wilson}.

However, it would be closer to the original spirit  of the RG to really
perform  the integrations over the short wavelength degrees  of freedom
step by step, i.e.\ by explicit computation of effective actions.

Of central interest in lattice quantum field theory is  the removal of
lattice artefacts, i.e., taking the continuum limit. Lattice theories on
the renormalized trajectory are free of lattice artefacts. Simulating
them gives direct access to  the continuum results. For a recent
discussion see, e.g. the work on  `perfect actions'
in~\cite{perfect1}-\cite{perfect6}.  The authors computed the perfect
action for the  2D $O(3)$ model (and, more recently, also for $SU(3)$
gauge theory~\cite{perfect6}). The perfect action is the classical
approximation to the renormalized trajectory  in the vicinity of the UV
fixed point. It is claimed that the perfect action is essentially free
of cutoff effects even at small correlation lengths.

In this paper we present an attempt to compute  the renormalized
trajectory of the 2D $O(3)$ model by Monte Carlo.
In contrast to previous attempts in this direction, we 
try to genuinely iterate RG steps. This means that, e.g.,
the second block spin step is based on a simulation 
of the effective action resulting from the first RG step. 
For a recent study of the same model that uses 
blocks of increasing size instead of a genuine iteration,
see~\cite{bock}.

The objective of our project is two-fold: First, it  is a feasibility
study on the problem of computing effective actions (with a reasonable
number of  couplings) by MC. We employ two different MC methods for the 
computation of effective couplings.  Our conclusion will be that the
business is expensive  with respect to computer and software
resources.  The second question addressed is  how the effective action
should be parameterized.  We study various ansaetze with 4 to 13
coupling  constants. 

It was observed already in previous  work~\cite{candemon} that it is
very difficult  to keep the correlation length correctly  scaled by the
RG transformations. Note that under a block spin step with block size
two the dimensionless correlation length should be exactly halved. In
the present work,  we computed for a subset of our  RG trajectories the
running  coupling of L\"uscher et al.~\cite{running}.  For small enough
couplings  three-loop perturbation theory for the $\beta$-function
together with the exactly known $\Lambda/m$-ratio  allows to estimate
the infinite volume correlation length $\xi$. We find that for our
actions $\xi$ decreases too quickly  with the renormalization steps.
Roughly speaking, this means that the speed in the  space of coupling
constants is too large.  However, more important than the speed along
the trajectory  is the right position of the trajectory in the space of
effective actions. 
We can expect the actions to be almost  free from lattice artefacts
only if they are close to the true RG  trajectory.
A sensitive indicator of scaling violations is the step  scaling function
introduced by L\"uscher et al.~\cite{running}.  In the absence of
scaling violations, the continuum limit should be well approximated
already at small $L$.  In this respect our results are disappointing. 
Our effective actions do not scale  much better than the  standard action
investigated in~\cite{running}.  The conclusion will be that most likely
more higher order operators  (of order four and possibly of order six) 
should be included in  the ansatz for the effective action, e.g. 
similar to what is put into account in  the perfect action
of~\cite{perfect1}.

This paper is organized as follows:  
In section~\ref{models} we define the models, set up  our notation and
describe the block spin transformation used.   Our methods to compute
effective actions are specified  in section~\ref{methods}. In
section~\ref{flows} we present and discuss  our results for the RG flows
in the various truncation  and parametrization schemes.
Section~\ref{xistepscaling} deals with our results for the running
coupling  constant and the step scaling functions.  Conclusions follow.

\section{Model Definition and Block Spin Transformation}
\label{models}

In this paper we investigate spin models on 2-dimensional 
square lattices with periodic boundary conditions.
The spins $\varphi_x$ have three real components and 
are constrained to have unit length.
The partition function reads 
\be
Z = \int D\varphi \, \delta(\varphi^2-1 ) \, 
\exp[-S(\varphi)] \, , 
\ee 
where the integration measure is 
\be\label{measure}
D \varphi \, \delta(\varphi^2-1) 
 = \prod_x \left( d^3 \varphi_x \, 
\delta(\varphi_x^2-1) \right) \, , 
\ee
and the action is assumed to be invariant 
under global $O(3)$ transformations (rotations). 
The so called standard action is written as  
\be\label{standardac}
S_{\rm st}(\varphi)= - \beta \sum_{<x,y>} \, \varphi_x \cdot \varphi_y \, , 
\ee 
where the sum runs over all (unordered) nearest neighbor pairs in 
the lattice. We shall write 
\be\label{action}
S(\varphi)= - \sum_i \beta_i S_i(\varphi) \, , 
\ee
with coupling constants $\beta_i$ and $O(3)$-invariant interaction
terms $S_i$. 
To be specific, we give here as examples interactions 
$S_i$ that will be used in the RG study below. The interaction terms 
are listed in a graphical notation in Table~\ref{tab1}.
With each graph there is associated an interaction 
$S_i$, $i=1..13$ as follows: 
Full circles connected by a line represent a scalar product
of the corresponding spins. Little empty circles are only there
to guide the eye. The full interaction term is obtained by summing
the object over all nonequivalent translations and reflections.
As an example, we have
\be\label{s7}
S_7 = \sum_{<x,y>} (\varphi_x \cdot \varphi_y)^2 \, .
\ee
We also consider a second kind of 
$O(3)$ invariant interactions associated with the same graphs:
Instead of assigning just the scalar product to a line between two
spins, choose one half of the angle squared between the two spins.
The corresponding interactions terms will be identified with $- S_i'$, 
and the couplings will be denoted by $\beta_i'$. 
We define
\be\label{ss1}
\theta_{xy} = 
\arccos(\varphi_x \cdot \varphi_y) \, .
\ee
Again an example: 
\be
S_1' = - \half \sum_{<x,y>} \theta_{xy}^2 \, . 
\ee
It is this latter type of expansion variable that was used 
in~\cite{perfect1}. In the appendix we present a little study on the
decimation block spin RG for the 1-dimensional $O(3)$ model. The study
indicates that the $S_i'$ have better convergence properties as
the number of operators is increased. Similar conclusions will be drawn
from our RG study for the 2D model.
 
Let us now turn to the block spin definition.
Given a spin configuration $\varphi$ on the original 
lattice, we determine the block spin configuration on a 
block lattice coarser by a scale factor $B=2$ as follows: 
\begin{itemize}
\item[(a)] Identify the set of sites with coordinates $x=(x_1,x_2)$, 
         where both $x_1$ and $x_2$ are even, with the 
         block lattice.  
\item[(b)] On the block lattice, define 
         $\psi_x = \varphi_x + \epsilon (\Delta \varphi)_x$.
         Here, $\Delta$ denotes the usual nearest neighbor 
         Laplacian on the fine lattice, defined through 
         $(\Delta \varphi)_x = \sum_{y.nn.x} (\varphi_y - \varphi_x)$.
\item[(c)] Normalize to unit length: $\phi_x= \psi_x/|\psi_x|$.
         The field $\phi$, considered as a function of the 
         block sites, is the block spin field. 
\end{itemize}
Our definition contains a free parameter $\epsilon$. We fixed this 
parameter by requiring that the fixed point effective action 
in 2D massless free field theory should be as local 
as possible. Note that step (c) of the block spin definition  
is to be omitted for unconstrained fields.
By numerical computation we found that 
for free fields $\epsilon=0.05$ is a good choice. 
For this particular choice the block spins embedded on the even
sites of the original lattice, are defined through 
\be 
\phi_x = 0.8 \varphi_x + 0.05 \sum_{y.nn.x} \varphi_y \, , 
\ee 
which for the nonlinear $\sigma$-model changes to 
\be 
\phi_x = 
\frac{0.8 \varphi_x + 0.05 \sum_{y.nn.x} \varphi_y}
     {|0.8 \varphi_x + 0.05 \sum_{y.nn.x} \varphi_y|} \, .
\ee 
In Table~\ref{laplace} we show a few of the matrix elements of the fixed
point Laplacian $\Delta^*$.  The decay of $\Delta^*(0,x)$ along the
lattice axes goes like $\sim \exp(-3.1 x)$. This can be compared with
the Gaussian block spin definition of Bell and Wilson \cite{bell-wilson}
employed in ref.~\cite{perfect1}. There the fixed point Laplacian (with
parameter $\kappa=2$) decays like $\sim \exp(-3.44 x)$. Let us conclude
this section with the remark that good locality  properties of the
massless free field fixed point give no guarantee for a local flow in
the $O(N)$ model at small $\beta$.  This question can only be answered
by numerical experiment  (see section~\ref{flows} below). 

\section{The Methods to Compute Effective Actions} 
\label{methods}

In this section we describe the two methods  that we used to compute
effective actions for the  2D $O(3)$ model.  For previous studies
concerned with this problem see, e.g., 
refs.~\cite{okawa,falcioni,candemon}.

\subsection{Schwinger Dyson Equations}

In this subsection we shall derive linear Schwinger Dyson (SD) 
equations for the coupling constants of $O(3)$ symmetric spin models  as
introduced in the previous section. 

Let $A$ be any function of the spin configuration $\varphi$,  where
$\varphi$ can be a field on any of the block levels.  We start from the
identity 
\be\label{basic}
\int D \varphi \, \delta(\varphi^2-1) \, 
\Lx_x   
\left\{ A(\varphi) \exp[-S(\varphi)] \right\} \,
= 0 \, . 
\ee 
Here, $\Lx_x$ denotes the infinitesimal generator of 
rotations, 
$$
\Lx_x^{\alpha} = \epsilon_{\alpha \beta \gamma} \, \varphi_x^{\beta}
\, \frac{\delta}{\delta \varphi_x^{\gamma}} \, , 
$$
where $\epsilon$ is the totally antisymmetric tensor with 
$\epsilon_{123}=1$.
\Eq{basic} expresses the rotational invariance of 
the measure defined in \eq{measure}. 
{}From \eq{basic} we derive  
\be\label{SD1}
\langle A \Lx_x S \rangle = \langle \Lx_x A \rangle \, . 
\ee 
We now assume that the action is given in the form \eq{action}. 
Let us choose observables $A_i$ as follows: 
\be\label{choice}
A_i = \Lx_y S_i \,  .
\ee 
Plugging this into \eq{SD1} we get a system of 
linear equations for the couplings $\beta_i$:
\be\label{SDeq}
\sum_j \left\langle \left( \Lx_y S_i \right) \cdot \left(\Lx_x S_j \right)
\right\rangle 
\, \beta_j = - \left\langle \Lx_y \Lx_x S_i \right\rangle \, . 
\ee 
Note that our choice of \eq{choice} makes sure that the observables
occurring in \eq{SDeq} have $O(3)$ symmetry and have thus 
nonvanishing expectation value (i.e., lead to nontrivial 
SD equations). 

In our actual implementation we used 
\eq{SDeq} with $x=y$. For this case let us 
perform the differentiations 
defined by the angular momentum operators $\Lx_x$. 
With the definitions 
\be
\frac{\delta S_i}{\delta \varphi_x^{\alpha}} \equiv \psi_x^{i,\alpha} \, , 
\ee
and 
\be
\frac{\delta^2 S_i}{\delta \varphi_x^{\alpha} \delta \varphi_x^{\gamma}}
 \equiv \xi_x^{i,\gamma\alpha} \, , 
\ee
we arrive at the SD equations for the couplings  
$\beta$, 
\be\label{sdeq}
  T \beta = R \, . 
\ee
The matrix $T$ is given by 
\be 
T_{ij} = \left\LL \psi_x^{i} \cdot \psi_x^{j} \right\RR 
   - \left\LL (\varphi_x \cdot \psi_x^{i})  
   (\varphi_x \cdot \psi_x^j) \right\RR \, , 
\ee
and the right hand side is a vector $R$, with components 
\be
R_i = 2 \left\LL \varphi_x \cdot \psi_x^{i}\right\RR 
    + \left\LL (\varphi_x,\xi_x^{i} \varphi_x)  \right\RR
    - \left\LL {\rm Tr}(\xi_x^{i}) \right\RR \, . 
\ee
\Eq{sdeq} allows for a straightforward determination of  effective
couplings using standard Monte Carlo methods.  One simulates the theory
on the  fine grid. The configurations are blocked according to the
chosen block spin rule, and the expectation  values giving the $T$
matrix and the vector $R$ are  measured.  Note that various blocking
schemes could be used simultaneously,  since the updating on the fine
lattice does not depend  on the blocking rule. 

\subsection{Canonical Demon Method}

The Canonical Demon method to compute effective coupling constants  was
introduced in~\cite{candemon}. The basic idea is to simulate a joint
system of block spins and canonical demons. The block couplings can then
be determined from the demon distribution. 

We assume again that the (effective) action can be written in the form
of \eq{action}.  An additional auxiliary system is introduced, called
demon system, that is governed by the demon action
\be
S_D(d) = \sum_{i=1}^n  \beta_{i} d_{i} \, ,
\ee
where the $\beta_{i}$ are the same as in the block spin action, the
$d_i$ are real numbers in the interval  $[0,d_{{\rm max},i}]$,  and $n$
denotes the number of demons. One introduces as many demons as
(effective) couplings  are in the ansatz for the action. Note that the
ranges covered by the demons may differ  from coupling to coupling. In
the following we shall consider the joint partition function
\be\label{jointZ}
 Z_{\rm joint} = \int D\varphi \, \delta(\varphi^2-1)
 \left( \prod_i \int_{0}^{d_{{\rm max},i}} {\rm d} \, d_i \right)
 \, \exp[-S(\varphi)-S_D(d)] \, .
\ee
The partition function factorizes in the partition function of the spin
system and the partition functions of the single demons. Hence we can
compute the demon expectation values $<d_i>$ as functions of the
(unknown) $\beta_i$ exactly. One gets 
\be
<d_i>  = \frac{1}{\beta_i}
                \left( 1 -
                \frac{\beta_i d_{{\rm max},i}}
                     {\exp(\beta_i d_{{\rm max},i})-1}  \right) \, .
\ee 
This relation can be numerically solved with respect to $\beta_i$.  The
idea of the method is to simulate the joint system specified 
by~\eq{jointZ} and to measure the demon expectation values. 

Details on the question how to  do the simulations of the joint system
(without knowing the block couplings $\beta_i$)   are given in
ref.~\cite{candemon}.

\section{Results for the RG Flows}
\label{flows}

We employed the methods described in the previous section to compute 
the RG flow of the 2D $O(3)$ model for several subsets of  the thirteen
interaction terms displayed in Table~\ref{tab1}.  An overview of our
`projects' is given in Table~\ref{tabproj}. 

All flows were started using the standard action defined 
in~\eq{standardac} with $\beta=2.5$.   Then a number of blocking steps
with block size $B=2$ were performed,  doing a genuine iteration. This
means that the result  of a single $B=2$ step was used as the input for
the next iteration.   These steps will in the following be called 
`first blocking'. On the same lattice, we always computed  also the
couplings for the second blocking level by  blocking the fine lattice
configurations twice. The results will be referred  to as `second
blocking'. Note that in case of an exact  control of the RG, the `first
blocking' result of the $i$th iteration  of the block spin
transformation should coincide with  the `second blocking' result of the
$(i-1)$th iteration.  Truncation of the number of couplings (which is
infinite in  the exact RG) and possible effects from using  finite
lattice sizes will lead to deviations from  this coincidence (see the
discussion below).

We typically performed 7 to 10 iterations, thus covering three  orders
of magnitude in the length scale. All calculations were done on lattices
of size $32 \times 32$, i.e.,  the block lattices for the `first
blocking' were of size  $16 \times 16$, for the `second blocking' of
size $8 \times 8$. These are not huge lattices. However, it is a 
central element of the renormalization group hypothesis that  the
computation of effective couplings is not affected  by strong finite
size effects. Indeed, the finite size effects are  expected to be
exponentially small in the lattice size also in the critical  regime. Of
course, this hypothesis has to be checked in  each particular case. We
did comparative runs on even  smaller lattices (of size $L=16$). The
changes in the effective  couplings were within the statistical
errors.

Using the Schwinger Dyson method, we computed the flow for two different
ansaetze of the effective action with four and ten couplings,
respectively. In both cases we used the `scalar product' interaction
terms $S_i$. The two projects are labelled by $SD_1$ and $SD_2$, cf.\
Table~\ref{tabproj}. The MC simulations were done by using alternatingly an
over-relaxation algorithm and a Metropolis updating of the spins on the
fine lattice. We performed two over-relaxation steps and three
Metropolis updates in change. The proposal for a new spin in the
over-relaxation update was given by a reflection of the old fields at
the sum of the spins at distance (1,0), (1,1)  and (2,0), weighted by
the corresponding couplings.  The acceptance rate for this step was
about 95\% in all simulations. The proposal for a new spin in the
Metropolis update was given by rotating the original spin by an angle
$\theta$ around the $x$, $y$ or $z$ axis, where each of the axes was
chosen with  probability $1/3$, and $\theta$ was chosen with uniform
probability from the interval $[-\theta_{max},\theta_{max}]$. We have
set $\theta_{max}$ such that the acceptance rate of the Metropolis
update was around 50-60\% in all runs. In order to obtain the effective
couplings, the linear equations were solved using the f04arf routine of
the NAG-library. The statistical errors quoted in the tables were
determined by using the jackknifing procedure.

Using the Canonical Demon method, we calculated the flow  for four
different ansaetze, with up to thirteen couplings.  One of these
ansaetze uses the `scalar product' interaction terms $S_i$, the three
others employ the `angle'  operators $S_i'$. The projects are labelled
by  $D_1$ to $D_4$, see Table~\ref{tabproj}.

Let us make a few remarks on the demon algorithm implementation.  In
order to update the demon/spin system we used a hybrid of a Metropolis
update and an over-relaxation algorithm (similar to those used in the SD
case).

For the demon-block spin updates a 2-hit Metropolis algorithm was used.
The proposals for a new spin were computed as described above. The
proposal was accepted, when each of the demons could take over its
corresponding change in the action. Five demon-block spin sweeps were
performed in a sequence. In order to avoid problems from correlated spin
configurations, without discarding too many of the generated configurations
entering the demon-block spin updates, we used 100 copies of
the demon set (cf.\ ref.~\cite{candemon}).

We want to give a rough account of the computational  costs of the
determination of the coupling constants:  For the projects $D_1$ and
$D_2$ the CPU time  on an Alpha AXP-3000/600 workstation  for a single
measurement was 10 seconds.  For $D_4$ (13 couplings) the costs were 19
seconds. 

One ``measurement" consisted of $100 \times (M+2\,O)$ for the
fine-lattice system. For each group of $(M + 2\,O)$, five
demon-spin-system sweeps were performed for the first and second-blocked
systems. Here, we have introduced for temporary use the  abbreviations
$M$ for Metropolis sweep and  $O$ for over-relaxation sweep.

Typically 7000-8000 measurements were performed for $D_3$ and $D_4$
which means that per RG step 1 or 2 days of CPU, respectively, where
required on our fastest machine.

For the SD projects we quote only the overall computational cost.  The
calculations were done on an IBM-RISC6000/590 work station. The CPU
needed for the $SD_1$ project was approximately 110 h,  whereas the
$SD_2$ calculations required $340$ hours. 

As examples we present our results for the projects 
$SD_1$, $D_2$ and $D_4$ in Tables~\ref{tabsd1}, \ref{tabd2},
\ref{tabd4}, and \ref{tabd4c}. 

Let us start the discussion of the coupling constant results  with a
comparison of the two methods employed.  The two projects $SD_1$ (see
Table~\ref{tabsd1}) and  $D_2$ (see Table~\ref{tabd2}) use the same set
of couplings, namely three spin-spin couplings with increasing distance,
and  one quartic coupling (for the precise definition, see again 
Table~\ref{tabproj}).  There is no reason to expect that for a truncated
ansatz of the effective action  the two methods should yield identical
results for  the coupling constants. Therefore the similarity of  the
two flows is quite remarkable.  

We want to comment here also on the efficiency of the  methods (in the
sense of precision in the effective couplings per unit of computer
time): Doing careful  runs (with four coupling constants) of the two
programs for the SD and the demon method, respectively,  we found that
the two methods are of comparable efficiency. However this statement
might not be generally true.  We found that in the demon system long
autocorrelations might appear when the number of couplings and demons
increases. On the other hand, the SD equations are definitely more
difficult to program, and furthermore, the  program is not so easily
adapted to different  parametrizations as the demon program.  For
example, it seemed to us very difficult  to implement a SD procedure for
the  ``angle'' parametrization. (All our  RG data for this
parametrization were obtained  with the demon method.)

Our second issue is the question of locality of the  effective actions.
In project $D_4$, we computed  the 2-spin interactions up to distance 3.
In Fig.~1 we show the results for $\ln(|\beta_i'|)$,  plotted as
function of the distance, for the three first  RG steps in the
truncation scheme of project $D_4$.  The graph shows that the couplings 
decay exponentially fast with distance, with  a decay length of order
0.2 to 0.3. 

It is of course of great importance for genuine RG iterations
to be feasible that the effective actions have good locality
properties. However, in addition, the number of local 
operators needed to parameterize the effective theory with
sufficient precision should also be reasonably small. 
We look at the ratios
$\beta_9/ \beta_7$ of the project 
$SD_2$ and also
$\beta_9'/ \beta_7'$ of the 
projects $D_3$ and $D_4$: 
\begin{eqnarray}
SD_2: & \beta_9/  \beta_7=  0.291  \\ 
D_3:  & \beta_9'/ \beta_7'= 0.017 \\ 
D_4:  & \beta_9'/ \beta_7'= 0.012 
\end{eqnarray}
These numbers are always for the first blocking step,  but stay of the
same order of magnitude all over the computed trajectory.  At least in
the case of the `scalar product' parametrization  of the $SD_2$ project,
our observation indicates that higher  order operators can not safely be
neglected.  In this respect the `angle parametrization' seems  superior.
See also the appendix, where in the case  of an exactly solved 1D $O(3)$
model the  two parametrizations are compared. 

As a consistency check of our RG-flows, one can compare the `first
blocking' result of the $i$th iteration  of the block spin
transformation with  the `second blocking' result of the $(i-1)$th
iteration. The reader is invited to have a careful look at  the tables. 
The differences observed look very small.\footnote{In the light of the
observations of section~\ref{xistepscaling}, however, the small
discrepancies could also be interpreted as a  warning!}

\section{Running Coupling Constant and Step Scaling Functions}
\label{xistepscaling}

In contrast to the previous discussion, where the lattice spacing was
set to 1, we shall in this section, where appropriate, use dimensionful
quantities.

In order to monitor the flow of relevant and marginal couplings under
scale-changes  so called phenomenological couplings have been
introduced.  Nightingale~\cite{Night} introduced the quantity 
\be
 \bar{g}^2 = m(L) L \; , 
\ee
where $m(L)$ is the mass gap  on a lattice with extension $L$ in spatial
direction and infinite extension in time direction.
He demonstrated at
the example of the 2D Ising model,  using the exact solution, how this
quantity can be used to determine the  critical temperature and the
critical exponent $\nu$. He also outlined the  relation of his finite
size scaling technique with Wilson's renormalization group. 

In the context of $O(N)$ nonlinear $\sigma$-models this coupling was 
first studied by M. L\"uscher et al.~\cite{running}.
The running coupling of ref.~\cite{running} is defined by
\be
\bar{g}^2 = \frac{2}{N-1} m(L) L \, . 
\ee
The normalization factor is chosen such that at tree-level  $\bar{g}^2$
is equal to the bare coupling of the theory. 

On a finite lattice $\bar{g}^2$
obviously depends on the bare 
action $S$ and the width of the lattice in lattice units $L/a$, 
\be
 \bar{g}^2(S(\beta) ,L/a) =  m(S(\beta),L/a) L \;. 
\ee
Assuming that the infinite volume correlation length $1/am$  is a monotonously
growing function of the parameter $\beta$ we can trivially  reparametrize
$\bar{g}^2(S(\beta) ,L/a)$. Then the finite size scaling hypothesis is that
$\bar{g}^2(a m(\infty) ,L/a)$ splits into a continuum part and a correction 
to scaling part which vanishes as $a\rightarrow 0$, 
\be
 \bar{g}^2(a m(\infty),L/a) = \bar{g}^2(m(\infty) L) + f(a) \, , 
\ee
where $\lim_{a\rightarrow 0} f(a) =0 $. The particular form of the 
corrections to scaling $f(a)$ depends on the action chosen.

For $L/a \approx \xi$ and larger it is straightforward, in a MC
simulation, to compute the coupling $\bar{g}^2$ as a function of the
lattice width (in physical units). One computes the  infinite volume
correlation length as a function of $\beta$ and then performs a
simulation for $L/a = \xi$ at the same $\beta$-values. 

In order to probe smaller length scales the phenomenological
renormalization group approach is used. One computes the change of the
coupling as  the length scale is changed.  On the lattice one defines
the step scaling function
\be
 \Sigma(s,\bar{g}^2,a) = \bar{g}^2(s L/a, \beta) \, . 
\ee
Following the finite size scaling hypothesis we get 
\be
 \Sigma(s,\bar{g}^2,a) = \sigma(s,\bar{g}^2) + f(a) \, , 
\ee
where again $\lim_{a\rightarrow 0} f(a) =0 $. 
The step scaling function is related to the $\beta$-function by
\be
\left.
\frac{ \partial \sigma(s,u)}{ \partial s} \right|_{s=1} = - \beta(u) \; .
\ee
 The $\beta$-function for the running coupling $\bar{g}^2$
 up to 3-loop order is  given by \cite{running}
\be
 \beta(\bar{g}^2) = 
     - \frac{N-2}{2 \pi} \bar{g}^4 - \frac{N-2}{(2 \pi)^2} \bar{g}^6
     -\frac{(N-1)(N-2)}{(2 \pi)^3} \bar{g}^8  ... \;\; .
\ee
The exact prediction for the mass gap given by
\cite{hasenfratz}
\be
\frac{m}{\Lambda_{\overline{MS}}} = \frac{8} {\mbox{e}}
\ee
for $N=3$, and
the conversion factor for the $\Lambda$ parameters
\be
 \Lambda = \frac{e^{-\Gamma'(1)}}{4\pi} \Lambda_{\overline{MS}}
\ee
given in ref.~\cite{running} allows us to give an estimate for the infinite
volume correlation length based on the measurement of the correlation length
on a  finite lattice.

In order to determine the mass on the lattices of width $L/a$ we simulated 
the theory with a multi-cluster algorithm specially adapted  to the actions
studied in this paper. The correlation functions were
measured using improved estimators. 

In our study we used $ \bar{g}^2$ for two purposes.
The first issue is the question of correct scaling of the correlation length
$\xi$. The second is the question of scaling violations
of the theory defined by our approximate effective actions.

We computed $\bar{g}^2$ for $L/a=4$ and $L/a=8$ in order to check 
how the correlation length changes under the approximate RG transformations
that we apply. Knowing $\bar{g}^2$ we can estimate the infinite volume 
correlation length using the exact result for 
$\frac{m}{\Lambda_{\overline{MS}}}$, the conversion factor for the 
$\Lambda$ parameters and the 3-loop result for the $\beta$-function.  This 
approach  is affected by two sources of systematical errors: \\
a) corrections to scaling due to the finiteness of  $L/a$, \\
b) contributions to the $\beta$-function  beyond 3-loop. 

In order to estimate the errors induced by a) we compare the results  based
on $L/a=4$ and $L/a=8$. 
Judging from the results of~\cite{running} one  expects that the errors 
stemming from b) are less the $10\%$ for $\bar{g}^2 < 1.0$. As an additional 
check we compared the  3-loop result with that of the 2-loop result.

The numerical results for the projects $SD_1$, $D_2$, $D_3$ and $D_4$
are summarized in the Tables~\ref{yxxx1}, \ref{xxx1}, \ref{xxx2}, \ref{xxx3},
and \ref{xxx4}, respectively.

Comparing with the 3-loop result for the correlation length we see that the 
coupling is running much too fast in the project $SD_1$. 

In the first step the correlation length is reduced by a factor of about
0.4 instead of $1/2$.  This mismatch is increased to a reduction factor
of about 0.3 in the $5^{th}$ step. Surprisingly in the  high temperature
limit given by the $10^{th}$ step the length rescaling  is almost
correct.  The results for $D_1$, which are not presented here, are
similar to those  of $SD_1$.  The length scales are sligthly better
reproduced in project $D_2$. Here we have a factor of about 0.43 for the
first step and about 0.36 in the $5^{th}$ step.

For $D_3$ the scaling of the correlation length is only slightly better
than for $D_2$. For the first step we have a factor of about $0.44$,
while it is about $0.37$ for the $5^{th}$ step.

Finally, for $D_4$ the scaling of $\xi$ is considerably improved.
For the first step we obtain a factor $0.5$,
while it is about $0.46$ for the $5^{th}$ step. 

We conclude that, at least for our starting action $\beta=2.5$, the
quality with which the scales are reproduced depends mainly on the
truncation of the  two-spin interaction of the action. The larger the
distances incorporated the more  accurate is the reproduction of the
scale. 

A comparison of the flow of $\bar{g}^2(L/a=4)$ for the various projects 
with the result from the 3-loop $\beta$-function  is given in
Fig.~2. After a few RG steps, the results  from the different truncation
schemes differ considerably from  each other. The $D_4$ data stay
closest to the perturbative  result. 

Following Hasenfratz and Niedermayer, we have computed 
$\Sigma(2,1.0595,a/L)$ in order to check for the continuum behavior
of the theory defined by our RG trajectories. We have made computations
for  $L/a=4,6$ and~8.  In order to obtain a one-parameter family of
actions we linearly interpolated the actions obtained from our
RG-iterations. This  procedure seems justified by the fact that this is
the exact behaviour at tree-level, and secondly the changes of the
action from one step  to the next are small.  In order to obtain bare
actions such that $\bar{g}^2=1.0595$ for a given $L/a$ we used an
iterative  method. First we simulated the theory for  two guesses of the
correct action with a moderate statistics.  From the result of these two
simulations we computed a next guess by linear interpolation.  Then we
performed a simulation with higher statistics at this value of the 
action.

We have performed this analysis for the projects $D_1$, $D_3$, the
second blocking  of $D_4$ and $SD_1$. The results are summarized in
Table~\ref{tabstep} and Fig.~3. The figure shows our results  together
with standard action results of ref.~\cite{running}. 
The continuum limit value estimated in ref.~\cite{running} is given by
the full line in the bottom of the plot, together with dashed lines that
indicate the estimated error. The corrections to scaling for $SD1$ and
$D_1$ are even larger than those of the  standard action~\cite{running}.
The corrections to scaling for $D_3$ are  about half of those of the
standard action for $L/a=8$. However the result is  far away from the
ideal situation of vanishing corrections to scaling. 

A little bit surprising is that the result for $D_4$ is worse than that
of $D_3$. The numbers are much the same as those of $D_1$.  This result
makes clear that a reasonable  scaling of the correlation length with the
MCRG iterations does not directly  relate with the corrections to
scaling. 

Let us quote roughly the computer time spent in the running coupling and
step scaling function computations. In the $SD_1$ project we needed 60~h
CPU on a SNI SC900 MIPS R8000 10 processor machine using one processor
for the computation of the running couplings and about 50 h on the same
machine for the step scaling function.  For the calculation of the
running coupling and the step-scaling function of the other projects we
used about 3 to 4 months of CPU time on modern workstations.

\section{Conclusions}

We conclude from our observations that certainly more  operators
containing higher powers of fields have  to be included in the ansaetze.
Our largest ansatz, used in the project $D_4$, contains  five couplings
of order four, and one coupling of order  six. An extension of our
ansaetze to a size comparable to the  perfect action of
ref.~\cite{perfect1} is of course possible, though  probably a lot of
work. It would nevertheless be interesting to undertake this enterprise,
in particular to better understand the reasons for the success of the
perfect action. 

\section*{Acknowledgment}
One of us (KP) would like to thank C.~Wieczerkowski for many 
interesting discussions.
APG would like to thank the Regionales Hochschulrechenzentrum
Kaiserslautern (RHRK) for support. 


\section*{Appendix: RG for the 1-Dimensional O(3)-Model}

In this appendix, we present a little study of an RG for the 
1D $O(3)$-model, which sheds some light on the 
question of parametrization of the effective action. 
We start from a model with partition function 
\be
Z = \int \prod_i \left[d^3 \varphi_i \, \delta(\varphi_i^2-1) \right]
\, \exp\left[ \beta \sum_i \varphi_i \cdot \varphi_{i+1} \right] \, . 
\ee 
A very simple RG transformation in this case is decimation: 
every second spin variable is integrated out, leaving you with 
an effective theory for the remaining spins. 
The result is again a model with nearest neighbor interaction. 
The effective action reads 
\be 
S_{\rm eff} = - \sum_i \ln \left\{ 
\frac{\sinh\left[ 2\beta  \cos \left( \half \theta_{i,i+1} \right)\right] } 
     { 2\beta \cos \left( \half \theta_{i,i+1} \right) } 
 \right\} \, , 
\ee 
where $\theta_{i,i+1}$ denotes the angle between 
the spins at site $i$ and site $i+1$. We have relabelled
the sites in such a way that neighboring block sites are 
named with subsequent integer numbers.
Let us now compare this exact result for the effective action with 
its approximations by Taylor series. We used the two different choices 
of expansion variables that were also used in the Monte Carlo 
studies of the 2D model reported in this paper, namely (``cos'') 
\be\label{eeqq1}
\ln \left\{
\frac{\sinh\left[ 2\beta  \cos \left( \half \theta \right)\right] }
     { 2\beta \cos \left( \half \theta \right) }
 \right\} = \sum_{k=0}^4 A_k [1-\cos(\theta)]^k + ... 
\ee 
and (``angle'')
\be\label{eeqq2}
\ln \left\{
\frac{\sinh\left[ 2\beta  \cos \left( \half \theta \right)\right] }
     { 2\beta \cos \left( \half \theta \right) }
 \right\} = \sum_{k=0}^4 B_k \theta^{2k} + ...  
\ee
In Fig.~4 we show the comparison of these two approximations  with the
exact result for $\beta=10$. The expansion was done  to the order
indicated in eqs.~(\ref{eeqq1}) and~(\ref{eeqq2}). Obviously, the
expansion in the angle squared is much better in the ``large field
region'',  i.e.\ where the angle between neighboring spins is large. 

We also computed the correlation lengths for the different 
approximations. 
The correlation length is given by 
\be
\xi = - \frac{1}{\ln(\lambda_1/\lambda_0)} \, , 
\ee 
where the $\lambda_i$ denote the eigenvalues of the transfer matrix, 
and $\lambda_0$ is the largest of these. 
In our case (as a consequence of the rotational invariance), the 
transfer matrix is a function of $\cos(\theta)\equiv x$ alone,
i.e.\ $T=T(x)$. 
It is not difficult to show that its eigenvalues are given by 
\be
\lambda_l = 2\pi \int_{-1}^1 dx \, T(x) \, P_l(x) \, , 
\ee 
where the $P_l$ denote the Legendre polynomials.
Using these formulae, we computed $\xi$ for the original theory, 
for the exact block theory, and for the two expansions
given in eqs.~(\ref{eeqq1}) and (\ref{eeqq2}). The correlation 
length from the exact block theory was (within the numerical
precision) always one half of the original correlation length 
that we denote by $\xi_0$ in the following.
The results collected in Table~\ref{tabaa1} clearly indicate that 
the ``angle'' expansion is superior to the ``cos''
expansion. We conclude this appendix with a final observation: 
The correlation length in the ``cos'' expansion 
converges from below to the right value, whereas 
in the ``angle'' case we have a convergence from 
above. This corresponds to the fact that in 
Fig.~4 the ``cos'' action does not suppress 
large angles sufficiently, while the ``angle''
action is a little too large in this region.

\newpage
\listoftables

\section*{Figure Captions}

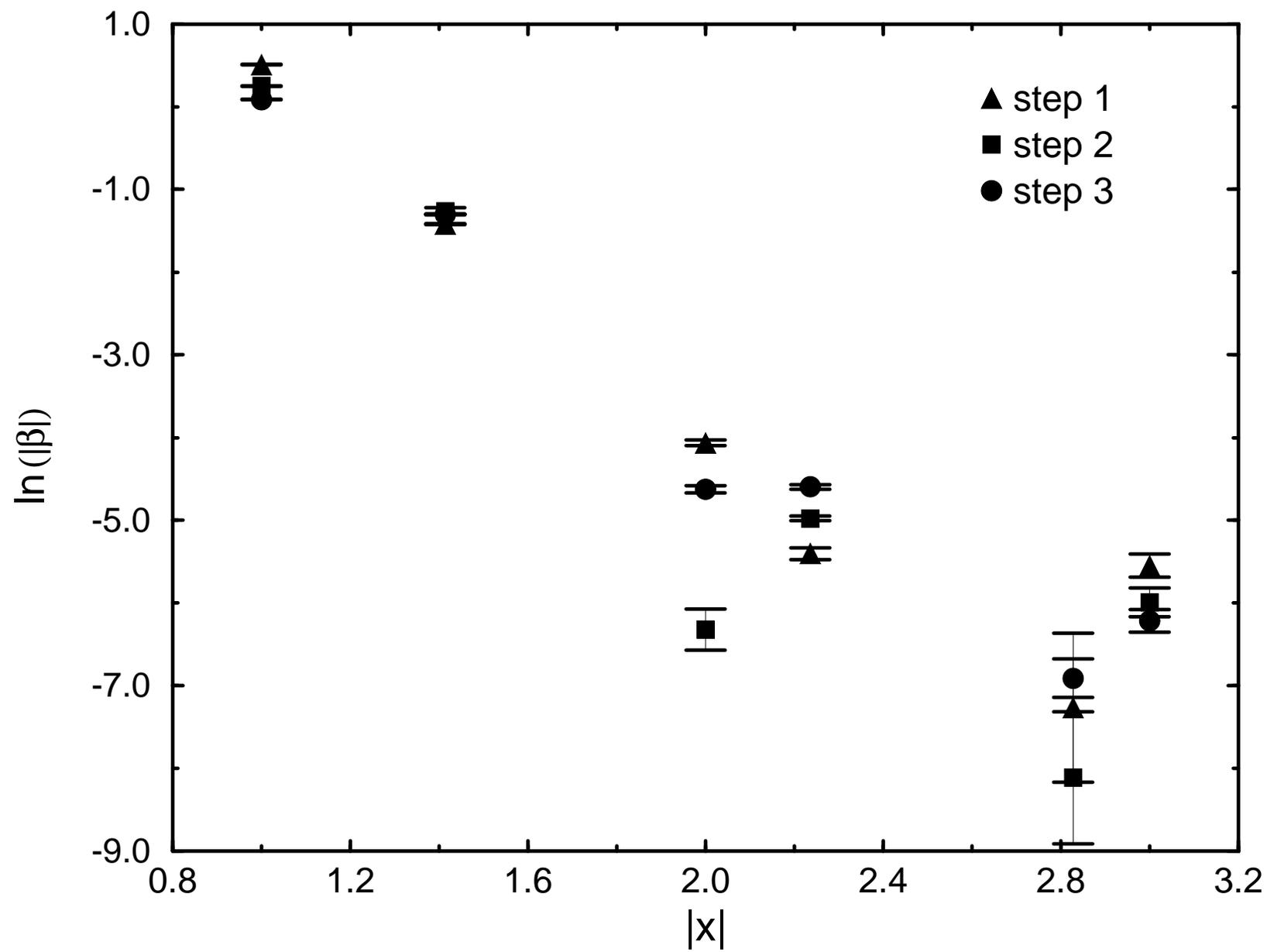
\noindent {\bf Fig.~1:} 
Our results for $\ln(|\beta_i'|)$,  plotted as
function of the distance, for the three first RG steps in the
truncation scheme $D_4$.

\vspace{0.5cm}

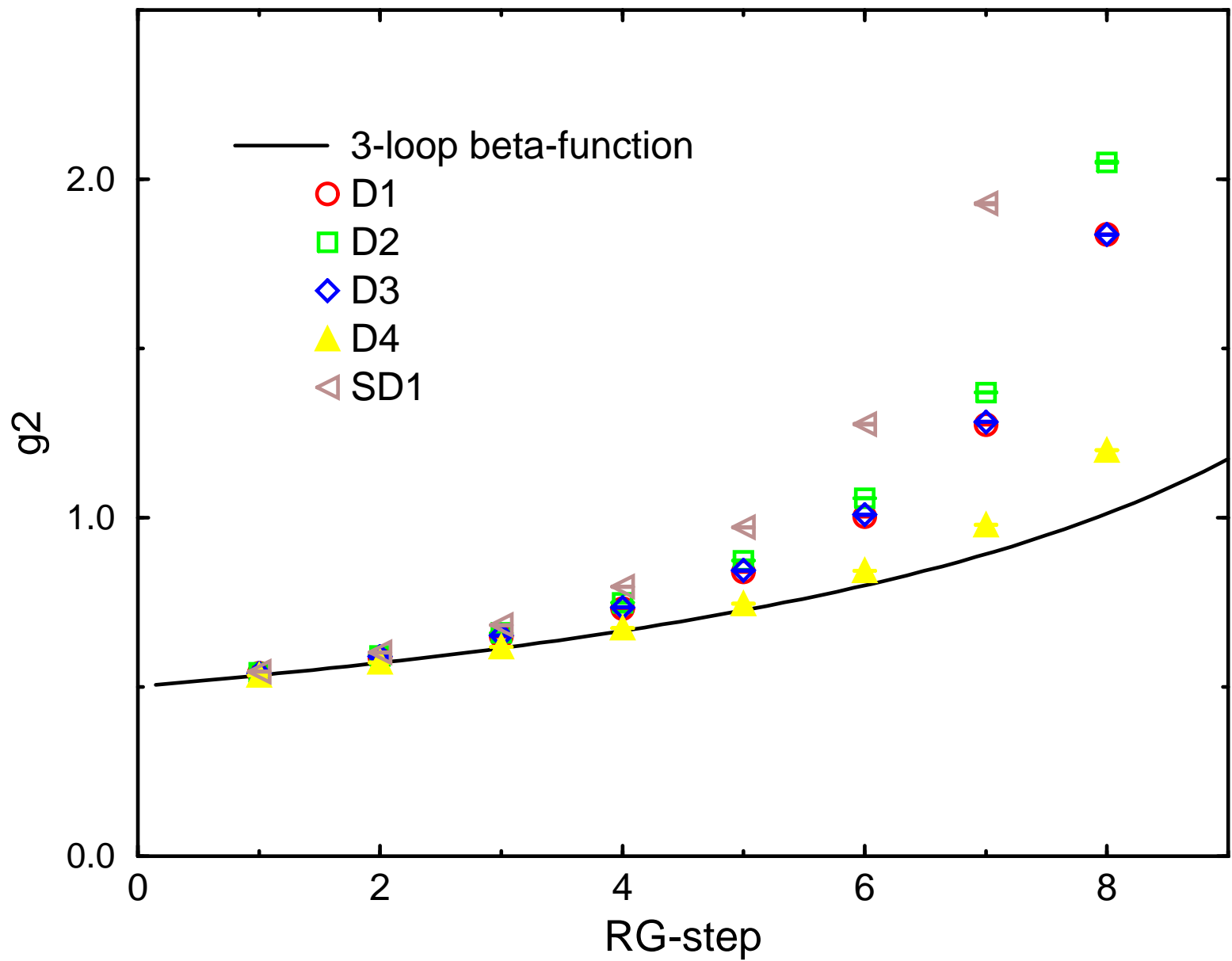
\noindent {\bf Fig.~2:}
Comparison of the flow of $\bar{g}^2(L/a=4)$ for the various projects 
with the result from the 3-loop $\beta$-function.

\vspace{0.5cm}

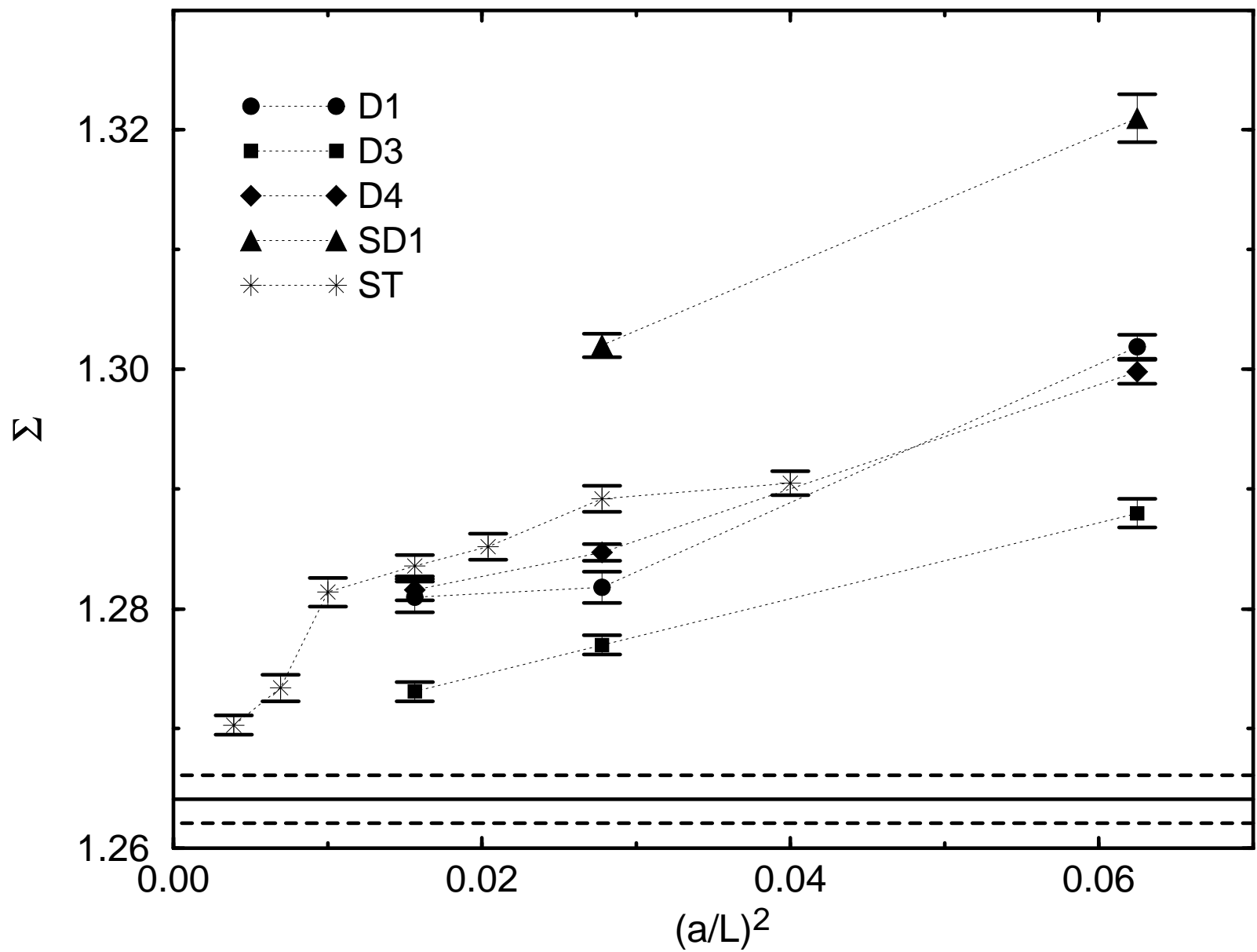
\noindent {\bf Fig.~3:}
Our results for the step scaling function for $\bar{g}^2=1.0595$
together with the results for the standard action of
ref.~\cite{running}.  The continuum limit value estimated in
ref.~\cite{running} is given by the full line in the bottom of the plot,
together with dashed lines that indicate the estimated error.

\vspace{0.5cm}

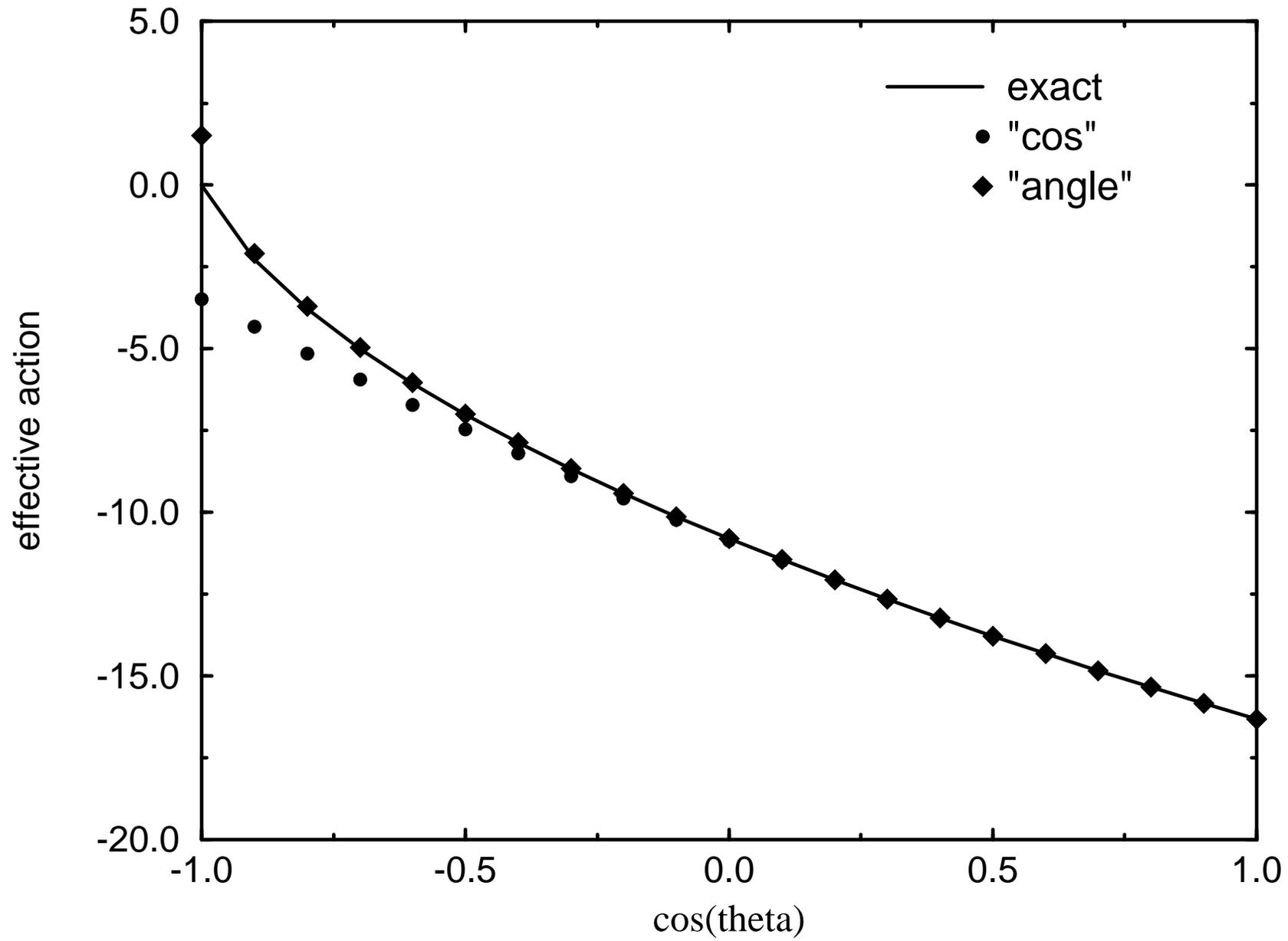
\noindent {\bf Fig.~4:}
Comparison of two approximations for the effective
action with the exact result in case of a decimation RG 
for the 1D $O(3)$ model. The two approximations are 
explained in the text. 


\newpage 

\begin{table}
\setlength{\unitlength}{0.8mm}
\begin{center}
\begin{tabular}{|r|c|c r|c|c r|c|}
\hline 
\# & type & ~~~~~~ & \# & type & ~~~~~~ & \# & type \\ \hline 

1 &
\begin{picture}(20,10)(0,3)    %
\put(5,5){\circle*{2}}         
\put(15,5){\circle*{2}}        %
\put(5,5){\line(1,0){10}}
\end{picture} & &

2 &
\begin{picture}(20,14)(0,5)    
\put(5,1){\circle*{2}}         
\put(15,11){\circle*{2}}       
\put(5,1){\line(1,1){10}}
\put(15,1){\circle{1}}  
\end{picture} & &

3 &
\begin{picture}(20,17)(0,5)   %
\put(0,5){\circle*{2}}        
\put(20,5){\circle*{2}}       %
\put(10,5){\circle{1}}       
\put(0,5){\line(1,0){20}}
\end{picture}  \\[1cm] 
\hline 
4 &
\begin{picture}(20,14)(0,5)
\put(0,1){\circle*{2}}    
\put(20,11){\circle*{2}}  
\put(0,1){\line(2,1){20}}
\put(10,1){\circle{1}}    
\put(20,1){\circle{1}}    
\end{picture} & & 

5 &
\begin{picture}(30,14)(0,5)    
\put(5,-4){\circle*{2}}         
\put(25,16){\circle*{2}}       
\put(5,-4){\line(1,1){20}}
\put(15,-4){\circle{1}}  
\put(25,-4){\circle{1}}  
\put(25,6){\circle{1}}  
\end{picture} & &  

6 &
\begin{picture}(30,20)(0,5)   %
\put(0,5){\circle*{2}}        
\put(30,5){\circle*{2}}       %
\put(10,5){\circle{1}}       
\put(20,5){\circle{1}}       
\put(0,5){\line(1,0){30}}
\end{picture} \\[1cm]
\hline
7 &
\begin{picture}(20,10)(0,3)    %
\put(5,5){\circle*{2}}         
\put(15,5){\circle*{2}}        %
\put(5,4.5){\line(1,0){10}}
\put(5,5.5){\line(1,0){10}}
\end{picture} & &

8 &
\begin{picture}(20,14)(0,5)    
\put(5,1){\circle*{2}}         
\put(15,11){\circle*{2}}       
\put(5,0.5){\line(1,1){10}}
\put(5,1.5){\line(1,1){10}}
\end{picture}  & & 

9 &
\begin{picture}(20,10)(0,3)    %
\put(5,5){\circle*{2}}         
\put(15,5){\circle*{2}}        
\put(5,4.3){\line(1,0){10}}
\put(5,5.0){\line(1,0){10}}
\put(5,5.7){\line(1,0){10}}
\end{picture}  \\[1cm]
\hline 
10 &
\begin{picture}(20,14)(0,5)    
\put(5,1){\circle*{2}}         
\put(15,1){\circle*{2}}        
\put(15,11){\circle*{2}}
\put(5,1){\line(1,0){10}}
\put(15,1){\line(0,1){10}}
\end{picture} & & 

11 &
\begin{picture}(20,14)(0,5)    
\put(5,1){\circle*{2}}         
\put(15,1){\circle*{2}}        
\put(15,11){\circle*{2}}
\put(5,1){\line(1,0){10}}
\put(5,1){\line(1,1){10}}
\end{picture}  & & 

12 &
\begin{picture}(20,14)(0,5)   
\put(5,1){\circle*{2}}        
\put(15,1){\circle*{2}}       
\put(15,11){\circle*{2}}      
\put(5,11){\circle*{2}}
\put(5,1){\line(1,1){10}}
\put(5,11){\line(1,-1){10}}
\end{picture} \\[1cm]
\hline  

13 &
\begin{picture}(20,14)(0,5)   
\put(5,1){\circle*{2}}        
\put(15,1){\circle*{2}}       
\put(15,11){\circle*{2}}
\put(5,11){\circle*{2}}
\put(5,1){\line(1,0){10}}
\put(5,11){\line(1,0){10}}
\put(-13,-9.5){\line(1,0){35.5}}
\end{picture}  \\[1cm]
\end{tabular}
  \parbox[t]{.85\textwidth}
  {
  \caption[Couplings used in the RG analysis]{\label{tab1}
  The interaction terms 
  used in the RG analysis. The symbolic notation is explained 
  in the text. This table defines the definition of 
  the couplings $\beta_i$ and $\beta_i'$. E.g., 
  $\beta_7$ multiplies
  the nearest neighbor interaction specified in \eq{s7},
  whereas the interaction associated with $\beta_1'$ is 
  given in \eq{ss1}}
  }
\end{center}
\end{table}
\begin{table}
\begin{center}
\begin{tabular}{|c|r||c|r|}
\hline 
$x$  & 
\mc{1}{|c||}{$\Delta^{*}(0,x)$}       &  $x$  & 
\mc{1}{|c|}{$\Delta^{*}(0,x)$}    \\[0.5mm] 
\hline
 &  &  &  \\[-4mm] 
(0,0)& $-0.3094 \cdot 10^{+01}$ & (1,3) & $0.2616 \cdot 10^{-03} $ \\
(0,1)& $ 0.5958 \cdot 10^{+00}$ & (2,3) & $0.6942 \cdot 10^{-05} $ \\
(1,1)& $ 0.1622 \cdot 10^{+00}$ & (0,4) & $0.1651 \cdot 10^{-04} $ \\
(0,2)& $ 0.3629 \cdot 10^{-02}$ & (1,4) & $0.1043 \cdot 10^{-04} $ \\
(1,2)& $ 0.5221 \cdot 10^{-02}$ & (3,3) & $0.4681 \cdot 10^{-06} $ \\
(2,2)& $ 0.1255 \cdot 10^{-03}$ & (2,4) & $0.2300 \cdot 10^{-06} $ \\
(0,3)& $ 0.6713 \cdot 10^{-03}$ & (0,5) & $0.1000 \cdot 10^{-05} $ \\
\hline 
\end{tabular}
  \parbox[t]{.85\textwidth}
  {
  \caption[Matrix elements of the fixed point Laplacian]
  {\label{laplace} A few of the matrix elements of the fixed point
  Laplacian computed for 2D massless free field theory 
  using the block spin transformation defined in section~\ref{models}
  with step (c) omitted}
  }
\end{center}
\end{table}
\begin{table}
\begin{center}
\begin{tabular}{|l|c|c|c|c|c|c|c|c|c|c|c|c|c|}
\hline
\mc{14}{|c|}{parametrization with spin products} \\ 
\hline 
          &
$\beta_1$ &
$\beta_2$ &
$\beta_3$ &
$\beta_4$ &
$\beta_5$ &
$\beta_6$ &
$\beta_7$ &
$\beta_8$ &
$\beta_9$ &
$\beta_{10}$ &
$\beta_{11}$ &
$\beta_{12}$ &
$\beta_{13}$ \\
\hline 
$SD_1$ & X & X & X &   &   &   & X &   &   &   &   &   &    \\
$SD_2$ & X & X & X &   &   &   & X & X & X & X & X & X & X  \\
$D_2$  & X & X & X &   &   &   & X &   &   &   &   &   &    \\
\hline
\mc{1}{}{} \\[0.5cm]
\hline
\mc{14}{|c|}{parametrization with angles} \\ 
\hline 
           &
$\beta_1'$ &
$\beta_2'$ &
$\beta_3'$ &
$\beta_4'$ &
$\beta_5'$ &
$\beta_6'$ &
$\beta_7'$ &
$\beta_8'$ &
$\beta_9'$ &
$\beta_{10}'$ &
$\beta_{11}'$ &
$\beta_{12}'$ &
$\beta_{13}'$ \\
\hline 
$D_1$ & X & X & X &   &   &   & X &   &   &   &   &   &    \\
$D_3$ & X & X & X &   &   &   & X & X & X & X & X & X & X  \\
$D_4$ & X & X & X & X & X & X & X & X & X & X & X & X & X  \\
\hline 
\end{tabular}
  \parbox[t]{.85\textwidth}
  {
  \caption[Parametrization schemes used in the RG analysis]
  {\label{tabproj} The parametrization
  schemes used (to be called `projects' in the text).
  The projects $D_i$ are based on the Canonical Demon Method,
  for the $SD_i$ schemes we used the Schwinger Dyson approach}
  }
\end{center}
\end{table}

\begin{table}
\begin{center}
\begin{tabular}{|r|l|l|l|l|}
\hline
\mc{5}{|c|}{first blocking} \\ 
\hline
step & \mc{1}{|c|}{$\beta_1$} 
& \mc{1}{|c|}{$\beta_2$} 
& \mc{1}{|c|}{$\beta_3$} 
& \mc{1}{|c|}{$\beta_7$} \\  
\hline
 1 & 2.05678(139) & 0.30911(63)&    -0.00948(51)& -0.21820(96) \\
 2 & 1.74818(127) & 0.36507(53)& \ph 0.01076(43)& -0.24499(98) \\
 3 & 1.52128(75)  & 0.35053(50)& \ph 0.02456(35)& -0.22273(74) \\
 4 & 1.33083(92)  & 0.31978(48)& \ph 0.03164(39)& -0.19238(80) \\
 5 & 1.15359(77)  & 0.28598(56)& \ph 0.03323(43)& -0.16114(71) \\
 6 & 0.97419(55)  & 0.25066(38)& \ph 0.03179(32)& -0.12594(58) \\
 7 & 0.78070(57)  & 0.21029(40)& \ph 0.02389(35)& -0.08231(71) \\
 8 & 0.54831(37)  & 0.14621(14)& \ph 0.01306(21)& -0.03530(57) \\
 9 & 0.27458(35)  & 0.06024(26)& \ph 0.00195(28)& -0.00638(44) \\
10 & 0.06898(28)  & 0.00702(23)&    -0.00022(24)& -0.00026(55) \\ 
\hline 
\mc{1}{}{}\\[0.3cm]
\hline
\mc{5}{|c|}{second blocking} \\ 
\hline
step & \mc{1}{|c|}{$\beta_1$} 
& \mc{1}{|c|}{$\beta_2$} 
& \mc{1}{|c|}{$\beta_3$} 
& \mc{1}{|c|}{$\beta_7$} \\  
\hline
 1 & 1.76762(233)&\ph 0.37516(136) &\ph 0.01742(132)&   -0.27063(170) \\
 2 & 1.55116(169)&\ph 0.35983(100) &\ph 0.03103(101)&   -0.25807(153) \\
 3 & 1.35857(152)&\ph 0.33648(75)  &\ph 0.03935(75) &   -0.23475(144) \\
 4 & 1.18231(160)&\ph 0.30381(96)  &\ph 0.04154(85) &   -0.20690(148) \\
 5 & 1.00058(146)&\ph 0.26719(89)  &\ph 0.04378(77) &   -0.17430(137) \\
 6 & 0.80285(94) &\ph 0.22935(68)  &\ph 0.03946(62) &   -0.12782(115) \\
 7 & 0.56880(69) &\ph 0.16776(73)  &\ph 0.02166(71) &   -0.06104(115) \\
 8 & 0.28129(67) &\ph 0.06984(57)  &\ph 0.00308(50) &   -0.01051(115) \\
 9 & 0.06965(52) &\ph 0.00805(52)  &   -0.00072(49) &   -0.00123(97)  \\
10 & 0.01229(55) &   -0.00056(54)  &\ph 0.00047(59) &\ph 0.00070(111) \\ 
\hline 
\end{tabular}
  \parbox[t]{.85\textwidth}
  {
  \caption[Results of project $SD_1$]
  {\label{tabsd1} The effective coupling
  constants of project $SD_1$}
  }
\end{center}
\end{table}

\begin{table}
\begin{center}
\begin{tabular}{|r|l|l|l|l|}
\hline
\mc{5}{|c|}{first blocking} \\ 
\hline
step & \mc{1}{|c|}{$\beta_1$} 
& \mc{1}{|c|}{$\beta_2$} 
& \mc{1}{|c|}{$\beta_3$} 
& \mc{1}{|c|}{$\beta_7$} \\  
\hline
 1&2.0779(17)& 0.3143(8) &    -0.0070(6) &-0.2427(13) \\
 2&1.7765(10)& 0.3758(6) & \ph 0.0144(4) &-0.2749(9)  \\
 3&1.5600(9) & 0.3653(5) & \ph 0.0297(4) &-0.2570(8)  \\
 4&1.3835(8) & 0.3387(4) & \ph 0.0382(3) &-0.2306(7)  \\
 5&1.2234(5) & 0.3096(4) & \ph 0.0428(3) &-0.2035(5)  \\
 6&1.0682(4) & 0.2808(3) & \ph 0.0450(3) &-0.1752(5)  \\
 7&0.9090(3) & 0.2515(3) & \ph 0.0419(2) &-0.1379(4)  \\
 8&0.7318(3) & 0.2094(2) & \ph 0.0316(2) &-0.0856(3)  \\
\hline 
\mc{1}{}{}\\[0.3cm]
\hline
\mc{5}{|c|}{second blocking} \\ 
\hline
step & \mc{1}{|c|}{$\beta_1$} 
& \mc{1}{|c|}{$\beta_2$} 
& \mc{1}{|c|}{$\beta_3$} 
& \mc{1}{|c|}{$\beta_7$} \\  
\hline
 1 &1.7862(27)&0.3811(17)&0.0242(14)& -0.2980(22) \\
 2 &1.5769(21)&0.3732(14)&0.0373(9) & -0.2896(18) \\
 3 &1.3982(20)&0.3506(10)&0.0476(9) & -0.2676(18) \\
 4 &1.2386(13)&0.3235(8) &0.0503(5) & -0.2472(12) \\
 5 &1.0801(12)&0.2950(7) &0.0553(6) & -0.2232(12) \\
 6 &0.9144(8) &0.2679(7) &0.0565(6) & -0.1890(10) \\
 7&0.7323(6) & 0.2287(5) &0.0491(4) & -0.1303(8)  \\
 8&0.5097(4) & 0.1566(4) &0.0243(4)  & -0.0521(6) \\
\hline 
\end{tabular}
  \parbox[t]{.85\textwidth}
  {
  \caption[Results of project $D_2$]
  {\label{tabd2} The effective coupling
  constants of project $D_2$}
  }
\end{center}
\end{table}

\begin{table}
\begin{center}
\begin{tabular}{|l|l|l|l|l|l|}
\hline
\mc{6}{|c|}{first blocking} \\ 
\hline
 & \mc{1}{|c|} {step 1} 
& \mc{1}{|c|} {step 2} 
& \mc{1}{|c|} {step 3} 
& \mc{1}{|c|} {step 4} 
& \mc{1}{|c|} {step 5} \\  
\hline
$\beta_{1}$ & \ph 1.6627(17) & \ph 1.2832(14)& \ph 1.0894(15) &\ph 0.9583(14) &\ph 0.8543(17) \\
$\beta_{2}$ &  \ph 0.2420(12) &\ph 0.2822(12)&\ph 0.2728(11) &\ph 0.2511(11) &\ph 0.2295(10) \\
$\beta_{3}$ &  -0.0172(6)  &\ph 0.0018(4) &\ph 0.0098(4)  &\ph 0.0140(4)  &\ph 0.0168(3)  \\
$\beta_{4}$ &  \ph 0.0045(3)  &\ph 0.0069(2) &\ph 0.0101(3)  &\ph 0.0118(3)  &\ph 0.0126(2)   \\
$\beta_{5}$ &  -0.0007(5)  &-0.0003(2) &\ph 0.0010(3)  &\ph 0.0019(3)  &\ph 0.0022(3)  \\
$\beta_{6}$ &  \ph 0.0039(5)  &\ph 0.0025(4) &\ph 0.0020(3)  &\ph 0.0021(3)  &\ph 0.0020(2)  \\
$\beta_{7}$ &  -0.2023(2)  &-0.1487(21)&-0.1268(20) &-0.1141(17) &-0.1010(16) \\
$\beta_{8}$ &  -0.0391(8)  &-0.0478(7) &-0.0452(6)  &-0.0430(6)  &-0.0388(4)  \\
$\beta_{9}$ &  \ph 0.0025(10) &-0.0015(8) &-0.0007(8)  &-0.0011(6)  &-0.0013(5)  \\
$\beta_{10}$ &  \ph 0.0122(11) &\ph 0.0290(9) &\ph 0.0340(7)  &\ph 0.0351(6)  &\ph 0.0310(5)  \\
$\beta_{11}$ &  \ph 0.0294(9)  &\ph 0.0393(8) &\ph 0.0371(6)  &\ph 0.0358(6)  &\ph 0.0319(4)  \\
$\beta_{12}$ &  \ph 0.0055(14) &\ph 0.0038(12)&\ph 0.0027(9)  &\ph 0.0012(6)  &\ph 0.0011(5)  \\
$\beta_{13}$ &  \ph 0.0279(19) &\ph 0.0328(17)&\ph 0.0300(13) &\ph 0.0275(10) &\ph 0.0234(8)  \\
\hline 
\mc{1}{}{}\\[0.3cm]
\hline
\mc{6}{|c|}{second blocking} \\ 
\hline
& \mc{1}{|c|} {step 1} 
& \mc{1}{|c|} {step 2} 
& \mc{1}{|c|} {step 3} 
& \mc{1}{|c|} {step 4} 
& \mc{1}{|c|} {step 5} \\  
\hline
$\beta_{1}$ &  \ph 1.2788(33) & \ph 1.0909(33)&\ph 0.9578(31) &\ph 0.8511(33) &\ph 0.7600(34) \\
$\beta_{2}$ &  \ph 0.2822(31) &\ph 0.2685(21)&\ph 0.2484(18) &\ph 0.2263(19) &\ph 0.2049(18) \\
$\beta_{3}$ &  \ph 0.0018(11) &\ph 0.0092(9) &\ph 0.0138(9)  &\ph 0.0167(9)  &\ph 0.0180(6)  \\
$\beta_{4}$ &  \ph 0.0083(6)  &\ph 0.0098(7) &\ph 0.0116(5)  &\ph 0.0125(6)  &\ph 0.0129(4)  \\
$\beta_{5}$ &  \ph 0.0004(13) &\ph 0.0022(8) &\ph 0.0034(4)  &\ph 0.0022(7)  &\ph 0.0025(5)  \\
$\beta_{6}$ &  \ph 0.0028(8)  &\ph 0.0019(8) &\ph 0.0018(7)  &\ph 0.0030(6)  &\ph 0.0023(6) \\
$\beta_{7}$ &  -0.1471(39) &-0.1237(36)&-0.1135(42) &-0.1044(31) &-0.0971(28) \\
$\beta_{8}$ &  -0.0456(14) &-0.0469(15)&-0.0417(10) &-0.0412(10) &-0.0375(8) \\
$\beta_{9}$ &  \ph 0.0004(15) &-0.0026(12)&-0.0017(14) &-0.0025(9)  &-0.0007(9) \\
$\beta_{10}$ &  \ph 0.0286(21) &\ph 0.0325(19)&\ph 0.0333(15) &\ph 0.0348(7)  &\ph 0.0318(12) \\
$\beta_{11}$ &  \ph 0.0374(17) &\ph 0.0397(18)&\ph 0.0364(12) &\ph 0.0347(8)  &\ph 0.0301(8)  \\
$\beta_{12}$ &  \ph 0.0026(30) &\ph 0.0075(18)&\ph 0.0004(22) &-0.0007(17) &\ph 0.0010(15) \\
$\beta_{13}$ &  \ph 0.0342(38) &\ph 0.0323(33)&\ph 0.0304(29) &\ph 0.0306(20) &\ph 0.0255(19) \\
\hline 
\end{tabular}
  \parbox[t]{.85\textwidth}
  {
  \caption[Results for the first 5 steps of project $D_4$]
  {\label{tabd4} The effective coupling
  constants for the first 5 steps of project $D_4$}
  }
\end{center}
\end{table}

\begin{table}
\begin{center}
\begin{tabular}{|l|l|l|l|}
\hline
\mc{4}{|c|}{first blocking} \\ 
\hline
 & \mc{1}{|c|} {step 6} 
& \mc{1}{|c|} {step 7} 
& \mc{1}{|c|} {step 8} \\  
\hline
$\beta_{1}$ & \ph 0.7583(14) &\ph 0.6695(15)&\ph 0.5808(14) \\
$\beta_{2}$ &  \ph 0.2100(9)  &\ph 0.1858(8) &\ph 0.1656(8)  \\
$\beta_{3}$ &  \ph 0.0171(3)  &\ph 0.0173(3) &\ph 0.0164(3)  \\
$\beta_{4}$ &  \ph 0.0126(2)  &\ph 0.0119(2) &\ph 0.0109(2)  \\
$\beta_{5}$ &  \ph 0.0020(3)  &\ph 0.0025(2) &\ph 0.0025(2)  \\
$\beta_{6}$ &  \ph 0.0024(3)  &\ph 0.0023(2) &\ph 0.0020(2)  \\
$\beta_{7}$ &  -0.0922(16) &-0.0825(12)&-0.0721(10) \\
$\beta_{8}$ &  -0.0369(4)  &-0.0330(4) &-0.0303(3)  \\
$\beta_{9}$ &  -0.0018(5)  &-0.0017(3) &-0.0021(3)  \\
$\beta_{10}$ &  \ph 0.0299(5)  &\ph 0.0262(4) &\ph 0.0235(3)  \\
$\beta_{11}$ &  \ph 0.0292(4)  &\ph 0.0264(4) &\ph 0.0229(3)  \\
$\beta_{12}$ &  \ph 0.0023(6)  &\ph 0.0027(4) &\ph 0.0027(4)  \\
$\beta_{13}$ &  \ph 0.0205(8)  &\ph 0.0171(7) &\ph 0.0150(5)  \\
\hline 
\mc{1}{}{}\\[0.3cm]
\hline
\mc{4}{|c|}{second blocking} \\ 
\hline
 & \mc{1}{|c|} {step 6} 
& \mc{1}{|c|} {step 7} 
& \mc{1}{|c|} {step 8} \\ 
\hline
$\beta_{1}$ &  \ph 0.6698(31) &\ph 0.5821(25)&\ph 0.5807(16) \\
$\beta_{2}$ &  \ph 0.1824(16) &\ph 0.1586(15)&\ph 0.1655(6)  \\
$\beta_{3}$ &  \ph 0.0168(8)  &\ph 0.0163(5) &\ph 0.0164(3)  \\
$\beta_{4}$ &  \ph 0.0129(5)  &\ph 0.0126(4) &\ph 0.0109(2)  \\
$\beta_{5}$ &  \ph 0.0030(4)  &\ph 0.0019(5) &\ph 0.0024(2)  \\
$\beta_{6}$ &  \ph 0.0026(6)  &\ph 0.0017(4) &\ph 0.0020(3)  \\
$\beta_{7}$ &  -0.0836(2)  &-0.0767(22)&-0.0718(11) \\
$\beta_{8}$ &  -0.0343(7)  &-0.0307(5) &-0.0302(3)  \\
$\beta_{9}$ &  -0.0032(7)  &-0.0019(6) &-0.0023(3)  \\
$\beta_{10}$ &  \ph 0.0302(10) &\ph 0.0261(7) &\ph 0.0237(4)  \\
$\beta_{11}$ &  \ph 0.0279(6)  &\ph 0.0248(6) &\ph 0.0227(3)  \\
$\beta_{12}$ &  \ph 0.0005(17) &\ph 0.0021(9) &\ph 0.0029(4)  \\
$\beta_{13}$ &  \ph 0.0246(16) &\ph 0.0195(13)&\ph 0.0153(5)  \\
\hline 
\end{tabular}
  \parbox[t]{.85\textwidth}
  {
  \caption[Results for steps 6 to 8 of project $D_4$]
  {\label{tabd4c} The effective coupling
  constants for the steps 6 to 8 of project $D_4$}
  }
\end{center}
\end{table}

\begin{table}
\small 
\begin{center}
\begin{tabular}{|c|l|l|l|}
\hline
\mc{1}{|c}{step} & 
\mc{1}{|c}{$\bar g^2(L/a=4)$} &
\mc{1}{|c}{$\xi_{\rm 2loop}$}&
\mc{1}{|c|}{$\xi_{\rm 3loop}$} \\
\hline
\za  1 &\za  0.5458(4)  & 1672.(22.)    &    1831.(15.)\\
\za  2 &\za  0.6026(5)  &\za 623.6(5.)  &\za  689.7(5.)\\
\za  3 &\za  0.6835(6)  &\za 205.9(1.5) &\za  231.(1.5) \\
\za  4 &\za  0.7967(6)  &\zb 65.0(4)    &\zb 74.4(4) \\
\za  5 &\za  0.9725(9)  &\zb 19.07(10)  &\zb 22.56(11) \\
\za  6 &\za  1.2771(20) &\zc 5.36(3)    &\zc 6.73(4) \\
\za  7 &\za  1.9286(28) &\zc 1.537(5)   &\zc 2.218(7)\\
\za  8 &\za  3.5367(94) &               &            \\
\za  9 &\za 7.2202(261)& &\\
 10 & 14.5455(529)& &\\
\hline 
\end{tabular}
  \parbox[t]{.85\textwidth}
  {
  \caption[Running coupling for project $SD_1$, first blocking]
  {\label{yxxx1} Running coupling squared $\bar g^2(L/a)$ for 
  project $SD_1$, first blocking}  
  }
\end{center}
\end{table}

\begin{table}
\small 
\begin{center}
\begin{tabular}{|c|l|l|l||l|l|l|}
\hline
\mc{1}{|c}{step} & 
\mc{1}{|c}{$\bar g^2(L/a=4)$} &
\mc{1}{|c}{$\xi_{\rm 2loop}$}&
\mc{1}{|c||}{$\xi_{\rm 3loop}$}&
\mc{1}{|c}{$\bar g^2(L/a=8)$} &
\mc{1}{|c}{$\xi_{\rm 2loop}$}&
\mc{1}{|c|}{$\xi_{\rm 3loop}$} \\ 
\hline
0 & 0.5025(3) &   4149.(28.) &   4511.(31.) & 0.5323(4)&4366.(35.)&4770.(38.) \\
1 & 0.5428(3) &   1772.(10.) &   1939.(11.) &       &          &   \\
2 & 0.5930(3) &\za 726.5(3.5)&\za 802.2(3.8)&       &          &   \\
3 & 0.6602(3) &\za 275.0(1.1)&\za 307.4(1.2)&       &          &    \\
4 & 0.7496(4) &\za 100.4(4)  &\za 114.0(4)   &       &          &    \\
5 & 0.8741(5) &\zb 35.47(13) &\zb 41.2(14)  &       &          &    \\
6 & 1.0582(6) &\zb 12.30(3)  &\zb 14.79(4)   &       &          &    \\
7 & 1.3708(9) &\zc 4.11(1)   &\zc 5.26(1)    &       &          &    \\
8 & 2.051(2)  &\zc 1.346(3)  &\zc 1.998(3)   &       &          &   \\
\hline 
\end{tabular}
  \parbox[t]{.85\textwidth}
  {
  \caption[Running coupling for project $D_2$, first blocking]
  {\label{xxx1} Running coupling squared $\bar g^2(L/a)$ for 
  project $D_2$, first blocking}  
  }
\end{center}
\end{table}

\begin{table}
\small 
\begin{center}
\begin{tabular}{|c|l|l|l||l|l|l|}
\hline
\mc{1}{|c}{step} & 
\mc{1}{|c}{$\bar g^2(L/a=4)$} &
\mc{1}{|c}{$\xi_{\rm 2loop}$}&
\mc{1}{|c||}{$\xi_{\rm 3loop}$}&
\mc{1}{|c}{$\bar g^2(L/a=8)$} &
\mc{1}{|c}{$\xi_{\rm 2loop}$}&
\mc{1}{|c|}{$\xi_{\rm 3loop}$} \\ 
\hline
1 & 0.5414(2) &1821.(7.)     & 1992.(8.)    &0.5794(5) &   1820.(15.) & 2005.(17.) \\ 
2 & 0.5880(3) &\za 788.3(3.9)&\za 869.6(4.2)&0.6357(5) &\za   764.(5.)&\za 851.(6.) \\
3 & 0.6495(3) &\za 316.6(1.3)&\za 353.0(1.4)&0.7084(6) &\za 308.9(2.1)&\za 348.2(2.3)\\
4 & 0.7317(4) &\za 120.3(5)  &\za  136.1(6) &0.8126(7) &\za 113.6(7)  &\za 130.5(7)  \\
5 & 0.8411(5) &\zb 45.26(17) &\zb 52.25(20) &0.9599(7) &\zb  40.98(17)&\zb  48.37(19)\\
6 & 1.0045(5) &\zb 16.03(4)  &\zb  19.08(5) &1.2043(13)&\zb  13.62(6) &\zb 16.849(7) \\
7 & 1.2759(8) &\zc 5.383(13) &\zc  6.755(16)&1.7193(15)&\zc  4.074(9) &\zc 5.609(11)  \\
8 & 1.8374(10)&\zc 1.721(2)  &\zc  2.433(3) &          &              &            \\
\hline
\end{tabular}
  \parbox[t]{.85\textwidth}
  {
  \caption[Running coupling for project $D_1$, first blocking]
  {\label{xxx2} 
  Running coupling $\bar g^2(L/a)$ for 
  project $D_1$, first blocking}  
  }
\end{center}
\end{table}

\begin{table}
\small 
\begin{center}
\begin{tabular}{|c|l|l|l|}
\hline
\mc{1}{|c}{step} & 
\mc{1}{|c}{$\bar g^2(L/a=4)$} &
\mc{1}{|c}{$\xi_{\rm 2loop}$}&
\mc{1}{|c|}{$\xi_{\rm 3loop}$} \\
\hline
1 & 0.5413(3) &  1824.(11.)    &    1996.(12.) \\
2 & 0.5908(4) & \za 752.9(4.9) & \za 831.0(5.3) \\
3 & 0.6532(3) & \za 301.4(1.2) & \za 336.3(1.3) \\
4 & 0.7358(7) & \za 115.3(8)   & \za 130.6(9) \\
5 & 0.8450(7) & \zb  43.92(23) & \zb 50.75(26) \\
6 & 1.0098(7) & \zb  15.60(6)  & \zb 18.58(6)  \\
7 & 1.2841(8) & \zc  5.250(13) & \zc  6.599(15)  \\
8 & 1.837(1)  & \zc   1.722(2) & \zc 2.434(3) \\
\hline 
\end{tabular}
  \parbox[t]{.85\textwidth}
  {
  \caption[Running coupling for project $D_3$, first blocking]
  {\label{xxx3} Running coupling squared $\bar g^2(L/a)$ for 
  project $D_3$, first blocking}  
  }
\end{center}
\end{table}

\begin{table}
\small 
\begin{center}
\begin{tabular}{|c|l|l|l||l|l|l|}
\hline
 &  \mc{3}{|c||}{first blocking} & \mc{3}{|c|}{second blocking} \\ 
\hline 
\mc{1}{|c}{step} & 
\mc{1}{|c}{$\bar g^2(L/a=4)$} &
\mc{1}{|c}{$\xi_{\rm 2loop}$}&
\mc{1}{|c||}{$\xi_{\rm 3loop}$}&
\mc{1}{|c}{$\bar g^2(L/a=8)$} &
\mc{1}{|c}{$\xi_{\rm 2loop}$}&
\mc{1}{|c|}{$\xi_{\rm 3loop}$} \\ 
\hline
1 & 0.5348(5) & 2076.(21.)   & 2269.(22.)    & 0.5680(3)& 1109.(6.)&1220.(6.)    \\
2 & 0.5734(5) & 1009.(9.)    & 1110.(9.)     & 0.6156(4)&\za 511.1(3.0)&\za 566.6(3.3) \\
3 & 0.6188(6) &\za 487.4(4.3)&\za 540.6(4.7) & 0.6715(3)&\za 238.4(9)  &\za 266.9(1.0)  \\
4 & 0.6748(6) &\za 228.8(1.7)&\za 256.4(1.9) & 0.7410(3)&\za 109.37(33)&\za 123.99(37) \\
5 & 0.7465(4) &\za 103.51(41)&\za 117.47(46) & 0.8375(4)&\zb 46.53(14) &\zb  53.69(16) \\
6 & 0.8437(5) &\zb 44.36(17) &\zb 51.24(19)  & 0.9595(5)&\zb 20.54(6) &\zb 24.24(7)    \\
7 & 0.9794(6) &\zb  18.35(6) &\zb 21.74(7)   & 1.1761(7)&\zc 7.537(20)&\zc 9.272(23)   \\
8 & 1.2001(7) &\zc  6.911(17)&\zc 8.543(20)  &          &          &             \\
\hline 
\mc{1}{|c}{step} & 
\mc{1}{|c}{$\bar g^2(L/a=8)$} &
\mc{1}{|c}{$\xi_{\rm 2loop}$}&
\mc{1}{|c||}{$\xi_{\rm 3loop}$}&
\mc{3}{c}{} \\ 
\cline{1-4}
  2  &  0.6184(4)  & \za  980.(6.) & 1088.(6.)     & \mc{3}{|c}{} \\ 
  4  &  0.7403(5)  & \za  220.(1.)   & \za 250.(1.)  & \mc{3}{|c}{} \\ 
  6  &  0.9621(5)  & \zb  40.46(12)     &\zb 47.78(14) & \mc{3}{|c}{} \\ 
  8  &  1.5172(11) & \zc 5.850(14)      &\zc  7.713(20) &\mc{3}{|c}{} \\ 
\cline{1-4}
\end{tabular}
  \parbox[t]{.85\textwidth}
  {
  \caption[Running coupling for project $D_4$]
  {\label{xxx4} Running coupling squared $\bar g^2(L/a)$ for 
  project $D_4$, first and second blocking}  
  }
\end{center}
\end{table}

\begin{table}
\begin{center}
\begin{tabular}{|c|c|c|c|c|}
\hline
$L$ & $D_1$& $D_3$ & $D_4$, 2nd blocking & $SD_1$  \\
\hline 
4 & 1.3019(10) &  1.2880(12) &  1.2998(10) & 1.321(2) \\ 
6 & 1.2818(13) &  1.2770(8) &  1.2847(7) & 1.302(2) \\
8 & 1.2810(13) &  1.2731(8) &  1.2816(9) &          \\ 
\hline
\end{tabular}
  \parbox[t]{.85\textwidth}
  {
  \caption[Step scaling function for $\bar g^2=1.0595$]
  {\label{tabstep} Step scaling function 
  for fixed $\bar g^2 = 1.0595$ }

  }
\end{center}
\end{table}

\begin{table}
\begin{center}
\begin{tabular}{|c|c|c|c|c|}
\hline 
$\beta$ &  $\xi_0$  &      order & $\xi_{\rm block}/\xi_0$, ``cos''  &
$\xi_{\rm block}/\xi_0$,  ``angle'' \\
\hline 
 2 &    1.609862  &   1  &    .436919 & .611268 \\
   &              &   2  &    .480009 & .512450 \\
   &              &   3  &    .492313 & .503656 \\ 
\hline 
 5 &    4.483701  &   1  &    .406884 & .544260 \\
   &              &   2  &    .470826 & .501922 \\
   &              &   3  &    .487790 & .501208 \\
\hline 
10 &    9.491220  &   1  &    .446066 & .519166 \\
   &              &   2  &    .490159 & .499997 \\
   &              &   3  &    .497356 & .500054 \\
   &              &   4  &    .499097 & .500012 \\
\hline 
20 &   19.495726  &   1  &    .474001 & .508926 \\
   &              &   2  &    .497404 & .499963 \\
   &              &   3  &    .499610 & .500003 \\
\hline 
\end{tabular}
  \parbox[t]{.85\textwidth}
  {
  \caption[Correlation lengths in the 1-dimensional $O(3)$-model]
  {\label{tabaa1}
  1-dimensional $O(3)$-model: Correlation length $\xi_0$ for the original 
  theory, and ratios of $\xi_{\rm block}/\xi_0$
  for the two approximate effective
  actions obtained by expanding the exact effective action 
  in powers of either $(1-\cos(\theta))$ or $\theta^2$.} 
  }
\end{center}
\end{table}


\begin{thebibliography} {999}

\bibitem{ma} S.-K. Ma, Phys.\ Rev.\ Lett.\ 37 (1976) 461.
\bibitem{swendsen} R.H. Swendsen, in: {\sl Real Space Renormalization, 
                   Topics in Current Physics}, 
                   Vol.\ 30, edited by Th.\ W.\ Burkhardt 
                   and J.M.J. van Leeuwen, Springer, Berlin, 1982. 
\bibitem{baillie} C.F. Baillie, R. Gupta, K.A. Hawick, and G.S. Pawley,
                  Phys.\ Rev.\ B 45 (1992) 10438, and references therein.
\bibitem{wilson} K.G. Wilson, in: {\sl Recent developments of gauge theories},
                 G. 'tHooft et al., eds., Plenum, New York, 1980;
                 S.H. Shenker and J. Tobochnik, Phys.\ Rev.\ B 22 (1980) 4462.
\bibitem{perfect1} P. Hasenfratz and F. Niedermayer,
                  Nucl.\ Phys.\ B 414 (1993) 785. 
\bibitem{perfect2} F. Farchioni, P. Hasenfratz, F. Niedermayer,
                   and A. Papa, \\
 Nucl.\ Phys.\ B 454 (1995) 638. 
\bibitem{perfect3} M. D'Elia, F. Farchioni, and A. Papa, 
Nucl.\ Phys.\ B 456 (1995) 313.  
\bibitem{perfect3a} 
F. Farchioni, P. Hasenfratz, F. Niedermayer, and
A. Papa, \\             
Nucl.\ Phys.\ B 454 (1995) 638. 
\bibitem{perfect4} M. Blatter, R. Burkhalter, P. Hasenfratz,
                   and F. Niedermayer, hep-lat/9508028.
\bibitem{perfect5}
T. DeGrand, A. Hasenfratz, P. Hasenfratz, and F. Niedermayer, \\ 
Nucl.\ Phys.\ B 454 (1995) 615. 
\bibitem{perfect6} 
T. DeGrand, A. Hasenfratz, P. Hasenfratz, and F. Niedermayer, \\ 
Nucl.\ Phys.\ B 454 (1995) 587. 

\bibitem{bock}
W. Bock and J. Kuti, Phys.\ Lett.\ B 367 (1996) 242.  
\bibitem{candemon} M. Hasenbusch, K. Pinn, and C. Wieczerkowski, 
                   Phys.\ Lett.\ B 338 (1994) 308. 
\bibitem{running} M. L\"uscher, P. Weisz, and U. Wolff, 
                  Nucl.\ Phys.\ B 359 (1991) 221. 
\bibitem{bell-wilson} T.L. Bell and K.G. Wilson, 
                      Phys.\ Rev.\ B 11 (1975) 3431. 
\bibitem{okawa} M. Okawa, Phys.\ Rev.\ Lett.\ 54 (1985) 963.   
\bibitem{falcioni} M. Falcioni, G. Martinelli, M.L. Paciello, 
                   G. Parisi, and B. Taglienti, \\ 
                   Nucl.\ Phys.\ B 265 (1986) 187.
\bibitem{Night} M.P. Nightingale, Physica A 83 (1976) 561.
\bibitem{hasenfratz} P. Hasenfratz, M. Maggiore, and F. Niedermayer, \\
                     Phys.\ Lett.\ B 245 (1990) 522.
\end{thebibliography}
\end{document}